\documentclass[twocolumn,prl,showpacs]{revtex4}

\usepackage{epsf,graphicx,amssymb,amsmath}

\begin{document}

\title{Experimental observation and characterization of the magnetorotational instability}

\author{Daniel R. Sisan, Nicol\'as Mujica, W. Andrew Tillotson, Yi-Min Huang, William
Dorland, Adil B. Hassam, Thomas M. Antonsen}
\author{Daniel P. Lathrop}
\altaffiliation{To whom correspondence should be addressed. Email:
dpl@complex.umd.edu}
\affiliation{Department of Physics,
IREAP, IPST, University of Maryland, College Park, MD 20742}

\date{\today}

\begin{abstract}
Differential rotation occurs in conducting flows in accretion disks
and planetary cores.  In such systems, the magnetorotational instability
can arise from coupling Lorentz and centrifugal forces to cause
large radial angular momentum fluxes.  We present the first experimental
observation of the magnetorotational instability.  Our system consists of
liquid sodium between differentially rotating spheres, with an imposed
coaxial magnetic field.  We characterize the observed patterns, dynamics and
torque increases, and establish that this instability can occur from a
hydrodynamic turbulent background.
\end{abstract}
\maketitle

Diverse astrophysical and planetary phenomena involve the close interplay of
rotation and magnetic field generation.
There is theoretical and computational evidence for a magnetorotational
instability
\cite{balbus 1991,balbus 1998} where magnetic fields
destabilize differential rotation in electrically conducting
flows \cite{velikhov,chandra 1960,chandra 1961}.
This instability is conceptually related to centrifugal instabilities
such as the development of Taylor vortices, but the addition of
Lorentz forces causes flows to be unstable even when centrifugally
stable.
An excellent description of the underlying process is given
in two reviews\cite{balbus 1998, bal2003}.
The instability
leads to radial outflow of angular momentum, which in an
astrophysical context implies the enhancement of the rate of
inward matter flow \cite{pringle}.
It also leads to a type of magnetohydrodynamic turbulence, different
from hydrodynamic turbulence by the forces in balance.  The
nature of and transition to magnetohydrodynamic turbulence is of much
recent interest \cite{moresco, krasnov, ponty}.
The magnetorotational instability is thought to affect
differentially rotating stars and planetary interiors \cite{balbus
1994,kitcha}, and the dynamics of accretion disks in protostellar
systems and around compact objects such as black holes
\cite{begelman}.
In a geophysical or planetary context, the magnetorotational instability
could serve as a mechanism for initial field growth leading to
dynamo states requiring a finite amplitude initiation.

All our current understanding and appreciation of the
magnetorotational instability derives from theory and
numerical simulation \cite{balbus 1991,balbus 1998,hawley}.
Until now, there have been no direct observations of these phenomena in
nature or the laboratory, although the suppression of
centrifugal instabilities has been documented \cite{donnelly}.
In nature, the development of this instability
occurs in contexts considerably more complicated than can be
simulated, a primary problem being the possible presence of small scale
hydrodynamic turbulence.
A central issue we address is the growth of the
magnetorotational instability within turbulent flows, which are
unavoidable in experimental liquid metal flows in the proper
parameter regime.

\begin{figure}[ht!]
\centerline{ \epsfxsize=80mm \epsffile{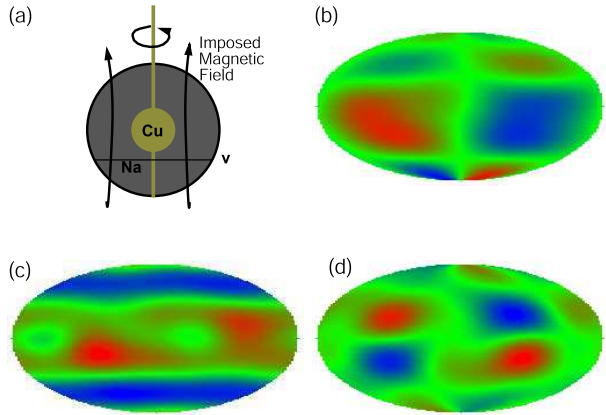} }
\caption{The spherical apparatus and characteristic induced
magnetic field instability patterns.  The device (a) consists of a
thin stainless spherical outer vessel, a rotating inner copper
sphere, and liquid sodium in between.  The line adjacent to {\bf
v} indicates the location of the velocity measurements. The
resulting dynamics change character under the influence of an
externally applied magnetic field coaxial with the rotation.  As
the external field increases, one sees states dominated by (b) a
rotating odd pattern with azimuthal wavenumber $m=1$ (O1) at
$R_m=26$ ($50$ Hz), $S=4.19$ ($84.6$ mT), (c) an even $m=0$
pattern (E0) at $R_m=10.4$ ($20$ Hz), $S=4.37$ ($88.1$ mT), and
(d) a rotating $m=1$ even pattern (E1) at $R_m=10.4$ ($20$ Hz),
$S=7.08$ ($14.29$ mT).  Red and blue indicate outward and inward
radial (cylindrical) induced magnetic field, and green indicates
the null values, in an equal area projection of the induced
magnetic field just outside the outer sphere. } \label{fig1}
\end{figure}

The experimental device (see Fig. \ref{fig1}) consists of sodium
\cite{note1} flowing between a rotating inner sphere (radius $a =
0.050$ m), and a stationary outer sphere. The inner sphere is made
of high conductivity copper, mounted on a $0.0125$ m radius rotating
non-magnetic stainless steel shaft which extends along the axis of
rotation.   The inner sphere rotates between $2.5$ to $50$
revolutions per second ($=\Omega_o/2\pi$).  The $0.010$ m thick
outer stationary vessel is a non-magnetic stainless steel shell
(radius $b = 0.15$ m). Due to its poor electrical conductivity, the
outer vessel is relatively passive in the magnetic field dynamics
\cite{peffley}. An external magnetic field, from $B_o = 0$ to
$0.2$ T, is applied co-axially using a pair of electromagnets.  We
observe the induced fields using an array of $30$ external Hall
probes \cite{note2} and the velocity using ultrasound Doppler
velocimetry \cite{takeda,note3}. The high electrical conductivity
of sodium, the highest of any liquid \cite{davidson}, allows
significant interactions between the fluid flow and the currents
causing induced magnetic fields, and allows an approach toward geophysically
realistic parameters.

In addition to the radius ratio $a/b$, three important
dimensionless numbers characterize our experiment. The magnetic
Prandtl number, $P_m = \nu/\eta= 8.8\times10^{-6}$, is the ratio
of kinematic viscosity $\nu$ to magnetic diffusivity $\eta$.  A
small value of $P_m$ is typical of liquid metals, planetary
interiors and other natural systems, but distinguishes them from
matter in accretion flows, for which $P_m \sim 1$.  The Lundquist
number, $S=B_o b / \eta (\rho\mu)^{1/2}$ (where $\rho$ is the
density and $\mu$ the magnetic permeability), is the ratio of the
Alfv\'en frequency to the resistive decay rate.  For $S \sim 1$,
system-size magnetic field oscillations within the sodium damp in
about one period; shorter wavelengths are damped more strongly.
The magnetic Reynolds number, $R_m = \Omega_o a b / \eta$,
characterizes the ability of the fluid motions to create induced
magnetic fields. Our experiments \cite{sisan} access the
relatively little-explored regime of parameter space $0<S<10$ and
$0<R_m<25$. Another important (but dependent) parameter is the
Reynolds number, $Re = R_m/P_m$. Because of the smallness of
$P_m$, the Reynolds numbers for our experiments are large (varying
between $1.7\times10^5$ and $3.4\times10^6$), implying
well-developed hydrodynamic turbulence.

\begin{figure}[t!]
\centerline{ \epsfxsize=85mm \epsffile{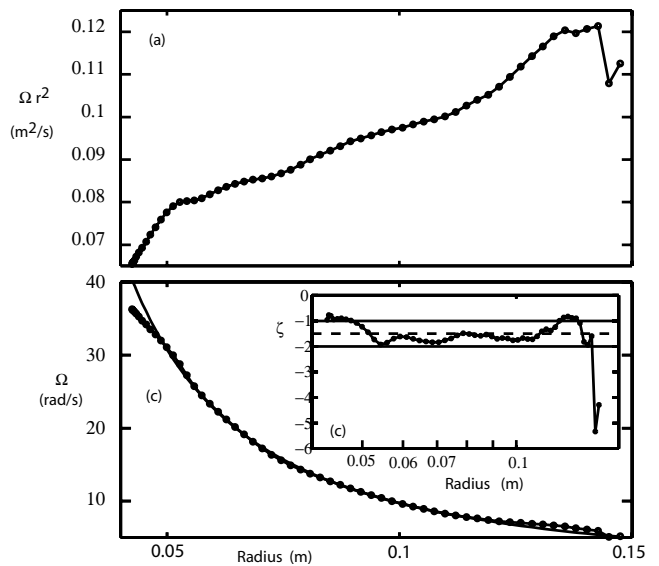} }
\caption{The zero field angular momentum and rotation rate
profiles for $\Omega_o /2\pi = 30$ Hz ($R_m=16$). From ultrasound
velocimetry measurements, the rising mean angular momentum density
(a) shows the system is stable to centrifugal instabilities except
for thin boundary layers. The rotation curve (b) decreases with
cylindrical radial distance from the center, a necessary condition
for the base state to be unstable to the magnetorotational
instability. The smooth curve is a Keplerian profile for
comparison. The inset (c) shows the velocity exponent $\zeta$ with
the Keplerian value $\zeta=-3/2$ indicated by a dashed line.}
\label{fig2}
\end{figure}

With no applied magnetic field (our base state), we have
examined the mean flow along a chord perpendicular to the axis of
rotation.  It has a profile with a velocity exponent $\zeta = \partial
\log \Omega(r) / \partial \log r$ in the range $\zeta \sim -1.4 $ to $-1.6$
except in thin boundary layers near the walls (see Fig.
\ref{fig2}). Here, $\Omega(r)$ is the rotation rate at a
cylindrical radius $r$, and would satisfy $\Omega(r) \sim r^\zeta$ were
$\zeta$ constant. For astrophysical rotation profiles
governed by Kepler's laws, $\zeta = -3/2$; that is, $\Omega^2r^3$
is constant. According to the hydrodynamic Rayleigh criterion,
flows are linearly stable for $\zeta > -2$. Nevertheless, we
observe $10-20\%$ turbulent velocity fluctuations in the base
state, likely generated in the boundary layers, which will be
fully described elsewhere.  Smaller turbulent fluctuations are
also observed in the magnetic field in the base state, due to
interactions between the fluid turbulence and the Earth's
relatively weak field, which is always present in the laboratory.
Profiles with $\zeta<0$ are predicted to be magnetorotationally unstable,
{\it assuming
a laminar base state}.  Precise stability boundaries have been
calculated for liquid metals ($P_m \sim 10^{-6}$) in a cylindrical
geometry \cite{ji,goodman}
and for $P_m \sim 1$ flows in spherical geometry \cite{kitcha}.
Aside from a simple rescaling with $P_m$,
the theoretical predictions are essentially identical: application
of an axial magnetic field of strength sufficient to overcome
resistivity will destabilize long-wavelength magnetorotationally driven
oscillations.

\begin{figure}[t!]
\centerline{ \epsfxsize=85mm \epsffile{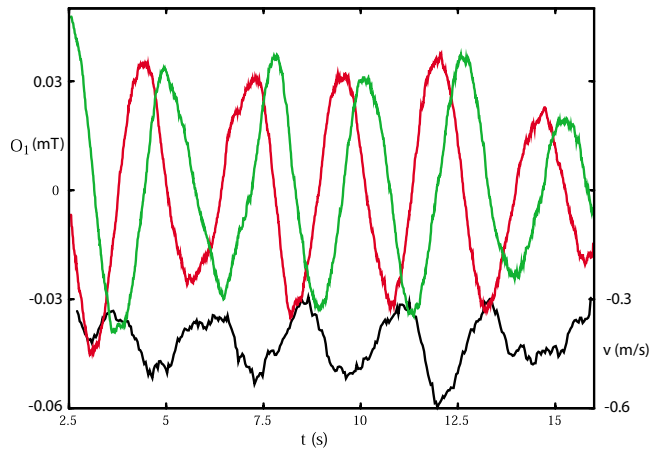} }
\caption{Coupled fluctuations in the velocity and magnetic fields.
Here the amplitude of the induced field for the two Gauss modes O1
(top curves) and the amplitude of a velocity component (lower
black curve) below the outer sphere substantiate that Lorentz
forces are key to this instability. These data were taken at
$\Omega_o/2\pi = 7.5$ Hz ($R_m = 4$); the velocity was measured
$0.115$ m from the outer wall.  Each mode with nonzero azimuthal
wave number has two components: when they oscillate $\pi/2$ out of
phase the pattern precesses.  The precession rate depends
(nonlinearly) on both applied field strength and rotation rate.}
\label{fig3}
\end{figure}

\begin{figure}[t!]
\centerline{ \epsfxsize=85mm \epsffile{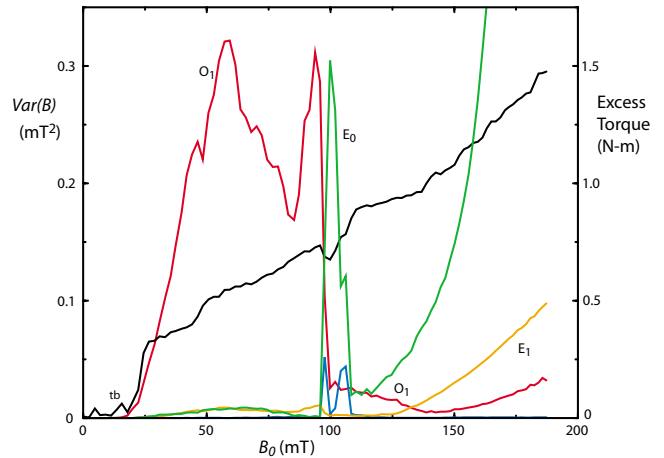} }
\caption{The torque increase and variance of induced field
coefficients as the applied field is varied.  We characterize the
saturated state by the mode with the largest variance.  These
measurements were taken for a fixed rotation rate of $30$ Hz
($R_m=16$). The field data reflect fluctuations in the
coefficients from their means, except the E0 state which shows
significant mean amplitude (squared average shown in green,
variance of fluctuations in blue), and does so at the same applied
field values where it shows ample fluctuations. At zero field, the
base state torque (including confounding errors due to seals) here
is $1.14$ N-m, the black curve shows the increase in torque above
this base value.} \label{fig4}
\end{figure}

Our primary observation is that for fixed rotation rate of the
inner sphere, above some threshold external magnetic field $B_c$,
we observe spontaneous excitation of oscillating magnetic
and velocity fields (Fig. \ref{fig3}). These take the form of a
rotating pattern with azimuthal wavenumber $m = 1$ (see Fig.
\ref{fig1}b). At instability onset the applied torque increases
(see Fig. \ref{fig4}), as increased amounts of angular momentum
are carried from the inner sphere to the fixed outer sphere. These
observations are evidence of the destabilization of the
magnetorotational instability.

\begin{figure}[t!]
\centerline{ \epsfxsize=85mm \epsffile{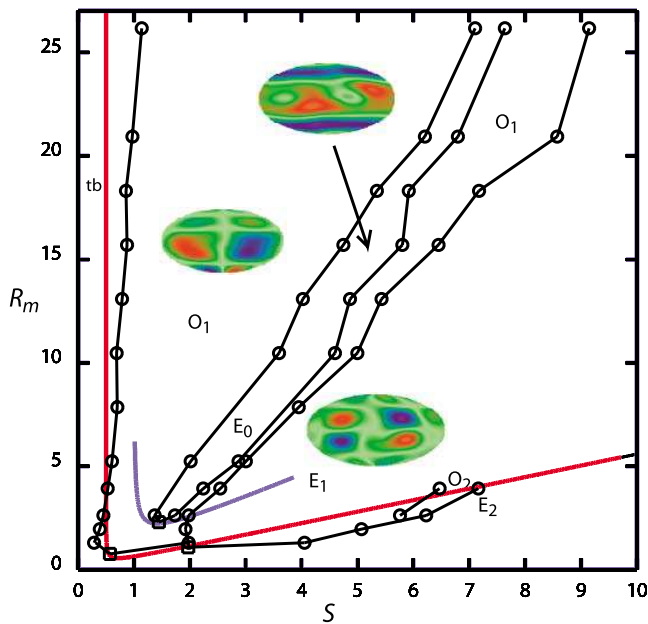} }
\caption{Phase diagram of saturated states.  Regions are defined
by the mode with the largest variance of fluctuations (see Fig.
\ref{fig4}).  Some secondary instabilities show hysteresis; these
data are for increasing $S$, for fixed $R_m$. The states have
regions associated with background turbulence (tb), mode O1
dominated, E0 dominated, followed by O1, E1, O2, and E2 modes. The
lowest $R_m$ and $S$ numbers for these states ($\square$) are
obtained by extrapolating the maximum saturated state amplitude
versus $R_m$ (a linear trend) to zero amplitude. Also shown are
theoretical stability boundaries for the longest wavelength (red)
and second longest wavelength (blue) instabilities, calculated
from the magnetorotational dispersion relation.} \label{fig5}
\end{figure}

The fields and flows consist of toroidal and poloidal components
\cite{bullard}. The azimuthal mode numbers $m$ and the parity
with respect to reflection through the origin characterize these
components. The primary observed magnetic field instability (Fig.
\ref{fig4}) is dominated by an $m=1$ poloidal perturbation that
has odd parity.  A toroidal magnetic field disturbance is also
likely, but our Hall array, being outside the sodium, can only
measure poloidal components. We denote the poloidal modes with the
notation E$m$ (azimuthal wavenumber $m$ with even parity) and O$m$
(azimuthal wavenumber $m$ with odd parity).  After our initial O1
instability, as the external magnetic field strength is further
increased, a number of other modes appear. These include saturated
states dominated by E0 (Fig. \ref{fig1}c), E1 (Fig. \ref{fig1}d),
O2, and E2. For applied fields many times $B_c$, the modes often show
aperiodic changes in pattern, rather than simple precession.  By
varying the rotation rate and external field independently, we
have navigated the ($R_m$,$S$) parameter plane and determined the
regions where these other modes dominate (Fig. \ref{fig5}).

Our primary instability consists of an $m=1$ pattern, in contrast
to the axisymmetric instabilities that dominate analogous
cylindrical situations.  Spherical calculations by Kitchatinov and
R\"udiger have shown $m=1$ instabilities in addition to
axisymmetric instabilities \cite{kitcha}. These differences  can be
understood by contrasting the different symmetries of the base
states.  In the cylindrical case, the base state is unchanged by
axial translations and rotations. Such situations generically show
instabilities to axially periodic patterns, and are known to do so
for the magnetorotational instability \cite{goodman,kim,noguchi}.
Our base state lacks the axial translation symmetry, but has
approximate rotational and reflectional symmetry.  In such
situations, instabilities involving rotating non-axisymmetric
patterns (Hopf bifurcations) are generic \cite{knobloch}. Indeed,
the amplitude of our disturbance shows the characteristic rise of
a Hopf bifurcation, for both the induced magnetic (see Fig.
\ref{fig4}) and velocity fields. However, the onset is made
imperfect by background turbulence and possibly geometric
imperfections. In a system similar to ours, Hollerbach and Skinner
\cite{hollerbach} found, numerically, rotating non-axisymmetric
patterns, though at much lower Reynolds number.

Also shown in Fig. \ref{fig5} are predicted magnetorotational stability
boundaries, as obtained from an inviscid dispersion
relation \cite{ji,goodman} adapted to the present configuration.
The two curves represent the stability boundaries of modes with $k
b = 1$ (red) and $k b = 2$ (blue) for wavenumber $k$, i.e. wavelengths
of one or one-half a circumference.
We take the
peak rotation rate $\Omega_p$ in the dispersion relation to be
an estimated fluid rotation just outside the inner
boundary layer $\Omega_p = \Omega_o/3$. The correlation between
the predicted stability boundaries and the experimental
observation of the magnetic field oscillations with progressively
more spatial structure is strong evidence that we have observed
the magnetorotational instability in our experiment.

These measurements, beyond being the first direct observation of
the magnetorotational instability in a physical system, have
important implications for our understanding of magnetohydrodynamic
instabilities. They establish that the magnetorotational
instability, and a significant increase in angular momentum
transport, occur in the presence of  pre-existing hydrodynamic
turbulence.   We quantify the nonlinear saturated amplitude of the angular
momentum transport and the patterns and
saturated values of magnetic field over a range of parameter
values not computationally accessible.
Finally, this geometry
may well be capable, at high $R_m$, of showing a dynamo instability. While
the dynamo instability is distinct from the magnetorotational
instability, it would be important to examine
a system capable of showing both.

\begin{acknowledgments}
Acknowledgments: We would like to acknowledge helpful discussions
with and assistance from: John Rodgers, Donald Martin, Yasushi Takeda,
James Stone, Eve Ostriker,
James Drake, Edward Ott and Rainer Hollerbach. This
work was supported by the National Science Foundation and the
Department of Energy of the United States of America.
\end{acknowledgments}


\begin{thebibliography}
\bibitem{note1} The experiments were carried out in liquid sodium
temperature controlled to $105\pm5^\circ$C.
\bibitem{note2} The induced field data from the 30 probe array were projected
onto the poloidal vector spherical harmonics of orders one through
four using least squares, giving 24 Gauss coefficients: one per
axisymmetric mode, two per nonzero m mode. Typical experiments
consisted of increasing the applied field from ambient to the
maximum external field (typically 0.20 T) in 2.2 mT increments for a
fixed rotation rate.
\bibitem{note3} The velocity measurements were obtained using pulsed
ultrasound Doppler velocimetry at 4 MHz. Oxide particles that
occur inevitably in our sodium scatter sound and track the flow
\cite{takeda}. We constructed an ultrasound transducer with a PTFE
surface coated by a thin film of silicon rubber in order to ensure
good wetting with the sodium.  The chord of velocity measurements
was 0.063 m below the equatorial plane and off axis by 0.043 m at
the nearest point.
\end{thebibliography}
\end{document}